# Thin-film based phase plates for transmission electron microscopy fabricated from metallic glasses

M. Dries[a,*], S. Hettler[a], T. Schulze[a], W. Send[a], E. Müller[a], R. Schneider[a], D. Gerthsen[a], Y. Luo[b] and K. Samwer[b]

[a] Laboratorium für Elektronenmikroskopie (LEM), Karlsruher Institut für Technologie (KIT), Engesserstraße 7, D-76131 Karlsruhe, Germany

[b] I. Physikalisches Institut, Universität Göttingen, Friedrich-Hund-Platz 1, D-37077 Göttingen, Germany

[*] Corresponding author. Tel.: +49 721 608 - 4 6489; Fax: +49 721 608 - 4 3721.
Email address: manuel.dries@kit.edu (M. Dries).

**Abstract**

Thin-film based phase plates are meanwhile a widespread tool to enhance the contrast of weak-phase objects in transmission electron microscopy (TEM). The thin film usually consists of amorphous carbon, which suffers from quick degeneration under the intense electron-beam illumination. Recent investigations have focused on the search for alternative materials with an improved material stability. This work presents thin-film based phase plates fabricated from metallic glass alloys, which are characterized by a high electrical conductivity and an amorphous structure. Thin films of the zirconium-based alloy $Zr_{65.0}Al_{7.5}Cu_{27.5}$ (ZAC) are prepared and their phase-shifting properties are tested. The ZAC-alloy film is investigated by different TEM techniques, which reveal a range of beneficial characteristics. Particularly favorable is the small probability for inelastic plasmon scattering, which is promising to improve the performance of thin-film based phase plates in phase-contrast TEM.

## 1. Introduction

Significant progress has been achieved in the past 15 years in the development of physical phase plates (PPs) for transmission electron microscopy (TEM). These developments are driven by weak or even vanishing contrast in TEM images of organic objects from biology, medicine and chemistry. Theoretical concepts for PPs have been available for decades. In his pioneering work [1], Boersch suggested to place a micrometer-sized Einzel lens in the back focal-plane (BFP) of the objective lens. The electrostatic field of the lens generates a phase shift of π/2 between unscattered and scattered electrons. Due to the demanding fabrication process, it took almost 60 years until the first electrostatic Boersch PP was realized [2-4]. A successful PP-variant, the thin-film based Zernike PP, was first studied and applied by Danev and Nagayama [5,6]. A thin amorphous carbon (aC)-film with a small hole in its center is positioned in the BFP of the objective lens. While the zero-order beam propagates through the central hole, a phase shift is imposed on the scattered electrons passing the aC-film. The phase shift

$$\varphi_{PP} = c U_{MIP} t \tag{1}$$

depends on the film thickness t, the mean inner potential (MIP) $U_{MIP}$ of the PP-material and the interaction constant c [7]. Impressive results have been achieved with aC-film based Zernike PPs, which include the tomographic reconstruction of phages and proteins [8,9]. An alternative approach are Hilbert PPs (HPPs), where one half of the BFP, excluding the zero-order beam, is covered by a thin film [10,11]. In this geometry, the film thickness has to be adjusted to a phase shift of π in order to obtain an image intensity, which is similar to that produced by a π/2-shift Zernike PP. The need for a doubling of the phase shift arises from the half-plane PP-design. If the mathematical expressions are derived, the phase shift has to be divided in symmetric and asymmetric parts, with which a factor of two is introduced. Another concept, the hole-free PP, applies a continuous film without any opening [12]. A relative phase shift between unscattered and scattered electrons is generated by the modification of the aC-film properties due to the intense illumination by the zero-order beam. This concept was recently pushed further by the so-called Volta PP [13]. After a defined sequence of preconditioning, the Volta PP reaches a steady state, in which a phase shift of π/2 is induced. In a recent review [14], numerous other PP-concepts are described and their potential as a phase-shifting device is evaluated.

Despite the impressive progress in the field of thin-film based PPs, aC-film based PPs still suffer from electrostatic charging, which causes a quick degeneration of the phase-shifting properties [15,16]. Hence, recent investigations have focused on the search for alternative materials with an improved material stability [17,18]. In this work, we present for the first time thin-film based PPs fabricated from metallic glass alloys. Metallic glasses are characterized by a high electrical conductivity and an amorphous structure. In particular, thin films of the zirconium-based alloy $Zr_{65.0}Al_{7.5}Cu_{27.5}$ (ZAC) were prepared and characterized by different TEM techniques. The ZAC-alloy was chosen for its high electrical conductivity of 0.47 $MSm^{-1}$ [19], which is about three orders of magnitude higher compared to aC-films with only 1 $kSm^{-1}$ [20]. Moreover, the ZAC-alloy remains in the amorphous state as long as the temperature stays below 437 °C [21], which prevents crystallization due to electron-beam induced heat-ing. The analytical TEM study of ZAC-films in this work shows that the ZAC-alloy is characterized by an extraordinary large inelastic mean free path, which reduces the coherence loss due to inelastic plasmon scattering. Proof of principle experiments carried out with ZAC-film based HPPs demonstrate that metallic glasses are promising materials to improve the performance of thin-film based PPs in phase-contrast TEM.

**2. Materials and Methods**

The ZAC-films were DC-magnetron-sputtered from an alloy target (*Goodfellow*) with the same stoichiometry as the intended composition of the ZAC-films. The basic and working pressures in the chamber amount to $1 \cdot 10^{-7}$ mbar and $5 \cdot 10^{-4}$ mbar (Ar), respectively. The structural isotropy of the metallic glass and the low sputtering pressure ensure that the film grows with a homogeneous thickness. Freshly cleaved mica sheets (*Plano – Part No. 54*) served as a substrate. A surfactant treatment of the mica substrates prior to the deposition reduces the adhesion of the ZAC-films and facilitates the floating process. The ZAC-films were floated off the mica substrates on a distilled water surface and deposited on Cu-grids (*Plano – Part No. G2150C, 150 mesh*). Subsequent plasma cleaning (*Binder Labortechnik – TPS 216*) for 7.5 minutes removes surfactant residua. Using a focused-ion-beam (FIB) system (*FEI Company – Strata 400 STEM*), rectangular windows with a size of 120 µm x 60 µm were structured into the ZAC-film. The Cu-grid was mounted on a customized objective aperture stripe (*Frey Precision*) and implemented in the BFP of a Philips CM 200 FEG/ST transmission electron microscope, which was operated at an acceleration voltage of 200 kV.

Two types of ZAC-film based HPPs were fabricated. The first type, denoted as ZAC-HPP, consists of a ZAC-film with 24 nm thickness. The second type, referred to as ZAC/aC-HPP, was fabricated in the course of the study, after it was found that oxide formation occurs at the surface of the ZAC-film. The ZAC/aC-HPP is composed of an 18 nm ZAC-film and coated with 4 nm aC on both sides. The aC-coating was deposited by electron-beam evaporation (*Kurt J. Lesker – PVD 75*) after completion of the PP-production process.

Plan-view and cross-section TEM samples were prepared to analyze the structural properties and the chemical composition of the ZAC-films. The cross-section TEM samples were prepared from ZAC-films on Si-substrates, which were deposited in the same sputtering process as the ZAC-films on mica-substrates. The preparation was carried out by FIB milling using the lift-out technique [22].

The structure of the ZAC-films was investigated by high-resolution TEM (HRTEM) and selected-area electron diffraction (SAED) using a 200 keV Philips CM 200 FEG/ST transmission electron microscope. The chemical composition of the ZAC-films was analyzed by energy-dispersive X-ray spectroscopy (EDXS) in the scanning TEM (STEM) mode in a FEI Titan$^3$ 80-300 operated at 300 kV. Composition quantification of the spectra was carried out by using the Cliff-Lorimer technique implemented in the FEI software "TEM imaging and analysis" (*TIA, Version: 4.6 build 1204*). Electron energy loss spectroscopy (EELS) in a 120 keV Zeiss 912 Omega was applied to determine the inelastic mean free path in the ZAC-alloy.

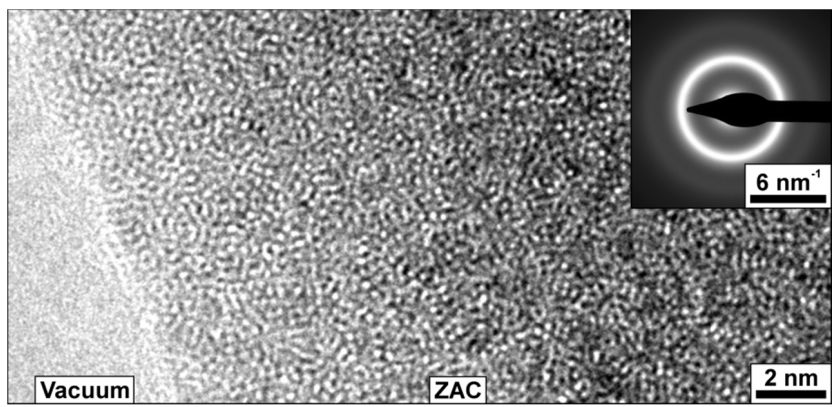

**Fig. 1:** Plan-view HRTEM image of a ZAC-film with inserted SAED pattern.

The phase shift induced by the ZAC-film was measured by off-axis electron holography in a FEI Titan[3] 80-300, which is equipped with a Möllenstedt biprism. The biprism consists of a gold-coated glass wire with a thickness of about 1 µm, which is positioned close to the first intermediate image plane. The holograms were taken with the biprism aligned parallel to an edge of the ZAC-film. Applying a voltage of 190 V to the biprism, the vacuum reference wave interferes with the object wave, which propagates through the metallic glass alloy. In the image plane, a hologram is formed, from which $\varphi_{PP}$ can be extracted. Details on the experimental procedure and the hologram evaluation are described in review articles [23,24].

## 3. Experimental Results

Fig. 1 presents a plan-view HRTEM image of a ZAC-film. The image does not show any indication of a long-range periodic order, which is confirmed by the SAED pattern shown in the inset. Only two diffuse rings, which reflect the distance between adjacent atoms, are observed due to short-range order in the amorphous material.

Fig. 2a shows a cross-section TEM image of a ZAC-film deposited on a thermally oxidized Si-substrate. An aC-layer was deposited on top of the ZAC-film to protect the sample during FIB preparation. Fig. 2a reveals a uniform film thickness, which is required to generate a homogeneous phase shift. In the upper region of the ZAC-film, a thin layer of about 5 nm thickness with a slightly brighter contrast arises from surface oxidation due to exposure in air. Fig. 2b presents the chemical composition of the ZAC-film derived from an EDXS line profile, which was taken along the white arrow in Fig. 2a. A substantial oxygen concentration is found in the ZAC-film, which is particularly high close to the ZAC/aC-interface. It is concluded that oxide formation occurs at the surface of the ZAC-film. The oxygen content in the central region of the ZAC-film is attributed to oxide formation at the surfaces of the cross-section TEM sample due to exposure in air. However, oxygen is not incorporated in the bulk material. The oxygen concentration is also enhanced at the $SiO_2$/ZAC-interface, which can be explained by the diffusion of oxygen from the $SiO_2$-layer into the ZAC-film. The analysis also reveals that the composition is quite inhomogeneous across the ZAC-film and the actual composition deviates from the intended one. In the middle of the ZAC-film, the composition is given by $Zr_{58}Al_{12}Cu_{30}$, whereas the nominal stoichiometry of the ZAC-alloy corresponds to $Zr_{65.0}Al_{7.5}Cu_{27.5}$. Nevertheless, the amorphous structure is preserved.

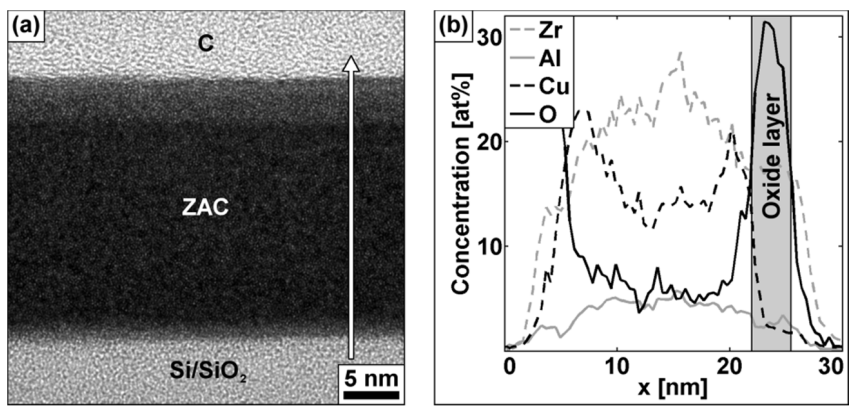

**Fig. 2:** (a) Bright-field cross-section TEM image of a ZAC-film deposited on a Si-substrate and (b) chemical composition derived from an EDXS line profile along the white arrow in (a).

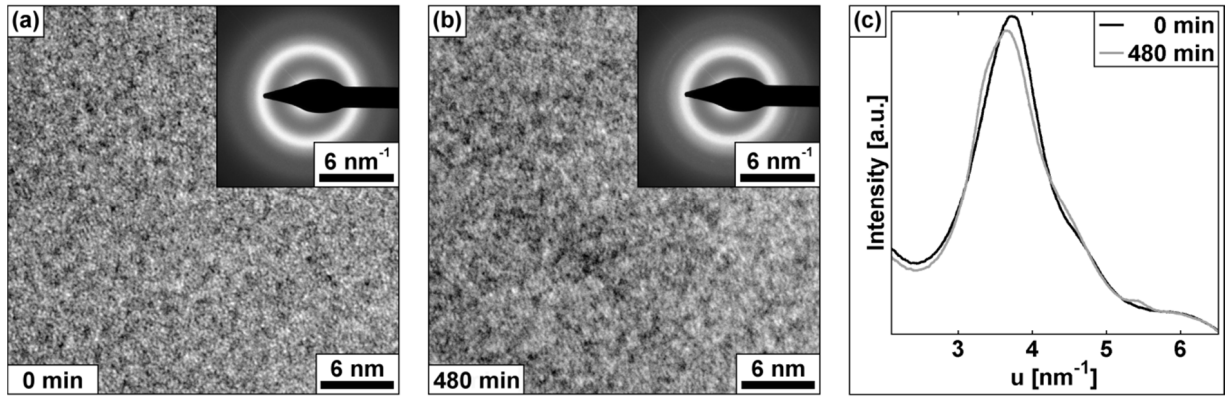

**Fig. 3:** Stability of the ZAC-alloy under electron-beam illumination. Bright-field TEM images and SAED patterns of a 24 nm ZAC-film acquired (a) at the beginning and (b) after 8 hours of electron illumination. (c) Intensity profiles obtained from azimuthally averaged SAED patterns.

The ZAC-alloy is expected to remain in the amorphous state, as long as the temperature stays below the crystallization temperature of 437°C. However, the temperature of TEM samples inserted in the object plane can reach a few hundred degrees due to electron-beam induced heating [25]. This can become more severe in the BFP of the objective lens, where the intensity is concentrated in the zero-order beam and Bragg-reflections in case of crystalline materials. Hence, the stability of the amorphous structure was tested by exposing a ZAC-film to a high electron dose. The ZAC-film was inserted in the object plane and illuminated for 8 hours with a total electron dose of $1.8 \cdot 10^7$ e/Å$^2$. In regular intervals, bright-field TEM images and SAED patterns were acquired to monitor structural changes of the ZAC-film. Figs. 3a,b show TEM images and inserted SAED patterns taken at the beginning and after 8 hours of illumination. The TEM image taken after 8 hours (Fig. 3b) does not show any obvious change of the amorphous structure if compared to the initial state (Fig. 3a). Only the SAED pattern indicates minor structural changes: After 8 hours of illumination, a weak Debye-ring can be recognized at 5.4 nm$^{-1}$ in the azimuthally averaged SAED patterns in Fig. 3c. Simultaneously, the intensity of the diffuse ring at 3.7 nm$^{-1}$, which is associated with the amorphous phase, decreases slightly and mar-ginally shifts to lower spatial frequencies. The appearance of the weak Debye-ring at 5.4 nm$^{-1}$ indicates a slight increase of short-range order, possibly induced by the rising temperature. A second, contrary effect of the intense electron-beam illumination is knock-on displacement resulting, e.g., in bond length changes between adjacent atoms. The slight shift of the diffuse ring towards lower spatial frequencies indicates an enlargement of the inter-atomic distance by knock-on damage.

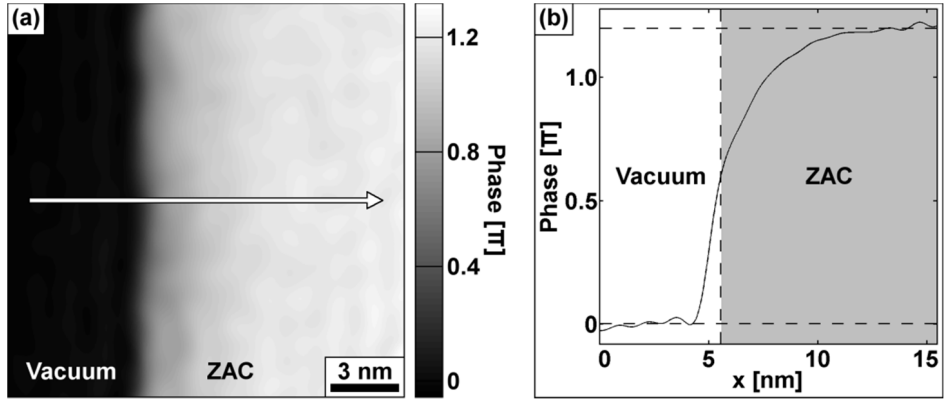

**Fig. 4:** Phase shift induced by the ZAC-HPP measured by off-axis electron holography. (a) Grayscale coded reconstructed phase and (b) phase line profile along the white arrow in (a).

The MIP can be approximately calculated by weighting the MIPs of the atomic constituents according to the material composition [26,27], which yields a MIP of 17.9 V for the ZAC-alloy. However, the weighted value can only be considered as a rough estimate of the true MIP because the actual composition deviates from the intended one as shown in Fig. 2b. According to Eq. (1), the true MIP can be derived if the phase shift and the film thickness are known. Hence, off-axis electron holography was applied to measure the phase shift induced by the ZAC-HPP by an independent technique. The grayscale coded reconstructed phase is shown in Fig. 4a, which contains a vacuum region on the left and the ZAC-HPP on the right. The phase shift is given in units of π and increases from 0 (black) to about 1.2 π (white). Fig. 4b shows a line profile along the white arrow in Fig. 4a. The phase shift rises steeply across the edge of the ZAC-HPP and reaches a plateau at about 1.2 π. Although the hologram was taken with 300 keV electrons, the phase shift in Fig. 4 is given for 200 keV for a better comparability with the following results. In combination with the film thickness, which is obtained by the investigation of cross-section TEM samples, the MIP can be derived according to Eq. (1). For a phase shift of 1.2 π and a film thickness of 24 nm, the MIP of the ZAC-alloy is given by 21.6 V. Table 1 summarizes the phase-shifting properties of the ZAC-HPP and the ZAC/aC-HPP. Based on the measured MIP of the ZAC-HPP, the phase-shifting behavior of the ZAC/aC-HPP can also be derived. The MIP of the aC-coating is given by 8.9 V, which was obtained by measuring the phase shift induced by an aC-film of known thickness by off-axis electron holography. While the ZAC-HPP induces a phase shift of 1.2 π, a phase shift of 1.1 π is expected for the ZAC/aC-HPP. This phase shift is composed of 0.9 π from the ZAC-film and 0.2 π from the aC-coating.

|  | ZAC-HPP | ZAC/aC-HPP |
|---|---|---|
| Structure (not to scale) | ZAC | aC / ZAC / aC |
| Thickness | 24 nm (ZAC) | 18 nm (ZAC) / 2 x 4 nm (aC) |
| MIP | 21.6 V (ZAC) | 21.6 V (ZAC) / 8.9 V (aC) |
| Phase shift | 1.2 π (ZAC) | 0.9 π (ZAC) / 2 x 0.1 π (aC) |

**Table 1:** Thickness, MIP and phase shift of the ZAC-HPP and the ZAC/aC-HPP.

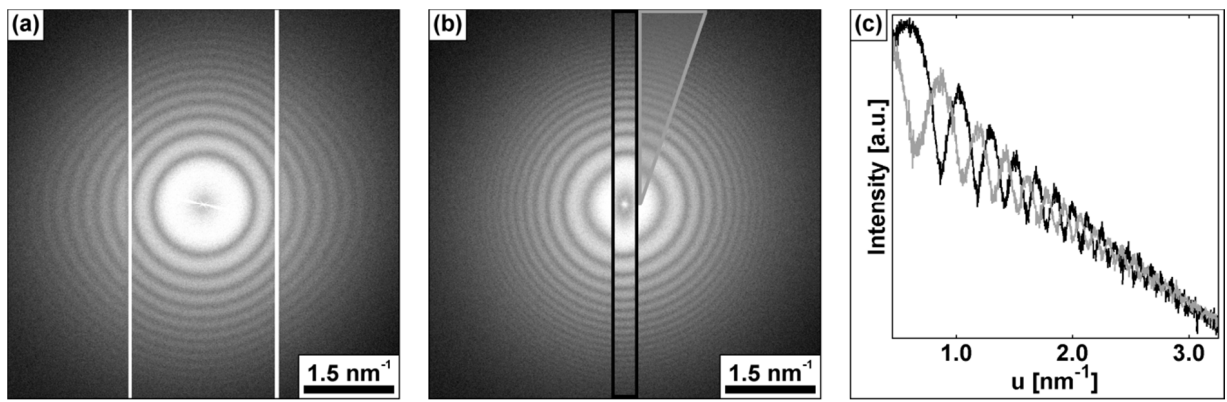

**Fig. 5:** (a) Application of the ZAC-HPP: Power spectrum of a phase-contrast TEM image of an aC test-object. Vertical white lines indicate the PP-edge. (b) Application of the ZAC/aC-HPP: Power spectrum of a phase-contrast TEM image of an aC test-object and (c) averaged intensity profiles obtained in the black and gray regions of Fig. 5b.

Oxidation of the surface of the ZAC-film leads to electrostatic charging upon electron exposure. This is illustrated in Fig. 5a, where the ZAC-HPP was used to image an aC test-object. The Fourier-transformed image intensity (power spectrum) can be subdivided in two regions: The inner region enclosed between the vertical white lines contains spatial frequencies not affected by the ZAC-film. Spatial frequencies in the outer regions are located above the cut-on frequency, which is given by the distance between the PP-edge and the zero-order beam. Electrostatic charging of the ZAC-HPP causes a slight distortion in the Thon-ring system. Moreover, a closer look at the Thon-rings reveals that the phase shift of 1.2 π anticipated by off-axis electron holography (cf. Table 1) is not reproduced. The phase shift of 1.2 π expected for the ZAC-HPP would result in a complementary behavior of the Thon-rings in the inner and outer regions of Fig. 5a. Instead, the Thon-rings are not shifted at the vertical white lines, which indicates a phase shift of 0 or 2π. Both values are incompatible with the measured film thickness of 24 nm and the MIP of 21.6 V. The apparently vanishing phase shift is a consequence of charging. Electrostatic fields generated by the charges cause an additional phase shift, which adds to the intended one. If the PP-edge is positioned closer to the zero-order beam, massive charging becomes visible by a strong distortion of the Thon-rings.

A substantial improvement is achieved for the ZAC/aC-HPP, where the thin aC-coating covers the oxidized surface of the ZAC-film and contaminants arising from PP-fabrication. This is demonstrated by Figs. 5b,c, which present the power spectrum of an aC test-object and averaged intensity profiles obtained in the black and gray regions of Fig. 5b. The gray region is confined to a narrow triangle in the upper half space because object drift affects the resolution in the horizontal direction. The zero-order beam was positioned in a distance of 0.8 µm to the PP-edge without any indication of charging as can be deduced from the absence of distortions in the Thon-ring system. The cut-on frequency is given by 0.2 nm$^{-1}$, which is a typical value for the CM 200 FEG/ST with its short focal length of 1.7 mm. Smaller cut-on frequencies can be realized at a larger focal length or in transmission electron microscopes equipped with a diffraction magnification unit. The phase shift of 1.1 π for the ZPP/aC-HPP (cf. Table 1) is indeed observed. Bright and dark Thon-rings connected at the vertical black lines demonstrate the expected complementary behavior above and below the cut-on frequency, which is reflected in the azimuthally averaged intensity profiles in Fig. 5c. It is pointed out that the attenuation of the Thon-rings towards high spatial frequencies is only slightly more pronounced above the cut-on frequency (Figs. 5b,c) indicating that the achievable information limit is not substantially impaired by inelastic electron scattering in the PP-material.

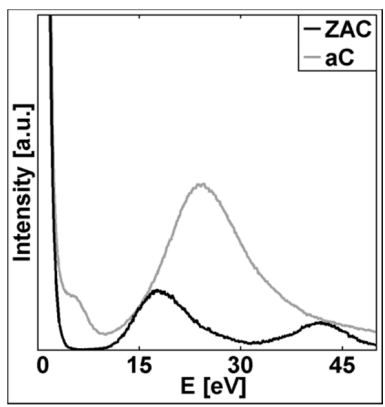

**Fig. 6:** EELS spectra of a 24 nm ZAC-film (black line) and a 49 nm aC-film (gray line).

Low-loss EELS spectra were acquired to study plasmon scattering, which is the dominant inelastic scattering process that limits resolution. Resolution is affected in three different ways. First, the inelastically scattered electrons are not able to interfere coherently, which reduces phase contrast. Second, the inelastically scattered electrons are slightly deflected, which yields a displacement of object information in the image plane and consequently a diffuse background and a reduction of the signal-to-noise ratio. Third, inelastic scattering reduces the electron energy, which leads to an additional blurring of the image by the effect of chromatic aberration. Fig. 6 presents low-loss EELS spectra of a 24 nm ZAC-film (black line) and a 49 nm aC-film (gray line) taken at an electron energy of 120 keV. Both films induce a phase shift close to π for an electron energy of 200 keV. Due to the different MIPs, the ZAC-film only needs to be about half as thick as the aC-film to achieve the same phase shift. The EELS spectra are normalized to the total number of counts, i.e., the total area below the curve. The EELS spectrum of the aC-film shows two plasmon signals with maxima at energy losses of 6 eV and 24 eV. The signal at 6 eV is associated with the π-π* transition and is regarded as a fingerprint for graphitic amorphous carbon with a high fraction of sp$^2$-hybridized carbon atoms [28,29]. The ZAC-plasmon peak is observed at an energy loss of 18 eV. The second ZAC-signal at 42 eV can be attributed to the Zr-N$_{2,3}$ ionization edge and is not related to plasmon scattering. The extraordinary low plasmon intensity of the ZAC-film compared to that of the aC-film demonstrates the small probability for plasmon scattering in the ZAC-alloy.

## 4. Discussion

The experimental results show that metallic glasses are promising materials to improve the performance of thin-film based PPs in phase-contrast TEM. The ZAC-alloy forms an amorphous structure, which is well preserved under intense electron-beam illumination. The oxidized surface of the ZAC-film was covered by a thin aC-coating to avoid electrostatic charging of the electrically insulating oxide layer. Phase-contrast TEM images of an aC test-object show indeed negligible charging of the ZAC/aC-HPP. Particularly favorable is the small probability for plasmon scattering in the ZAC-alloy. Even compared to aC, which is believed to be the best material in this respect, inelastic scattering in the ZAC-alloy is extraordinary low as demonstrated by the low plasmon intensity of the ZAC-film in Fig. 6.

Below, we discuss the effects of elastic and inelastic electron scattering in the PP-material. Inelastic scattering limits the transfer of high spatial frequencies and reduces the attainable resolution given by the information limit. However, elastic scattering cannot be neglected a priori because it modifies the direction of propagation. Elastically scattered electrons are deflected by the scattering angle, which causes a displacement of object information in the image plane [18]. In the following, the probability for electron scattering in ZAC- and aC-films will be compared. Film thicknesses of 49 nm and 24 nm are assumed for the aC- and ZAC-films, which yields the same phase shift close to π in these materials for an electron energy of 200 keV.

Elastic and inelastic scattering is assessed by the corresponding mean free path values ($\lambda_{elastic}$, $\lambda_{inelastic}$), which are proportional to $1/\sigma$ with $\sigma$ being either the total elastic or inelastic scattering cross-section. The total elastic scattering cross-section can be obtained by the Wentzel screening model [30]. Table 2 gives calculated values for $\lambda_{elastic}$ in aC and the ZAC-alloy at different electron energies of 120 and 200 keV. The elastic mean free path is considerably shorter in the ZAC-alloy than in the aC-film with its low material density. The effect of a short elastic mean free path is partially compensated by the higher MIP of the ZAC-alloy, which requires only 41 % of the aC-film thickness to achieve the same phase shift. The ratio between the film thickness t and $\lambda_{elastic}$ is of particular importance because

| Material | t [nm] | Electron Energy [keV] | $\lambda_{elastic}$ (calculated) [nm] | $P_{elastic}$ [%] | $\lambda_{inelastic}$ (calculated) [nm] | $\lambda_{inelastic}$ (measured) [nm] | $P_{inelastic}$ [%] |
|---|---|---|---|---|---|---|---|
| aC | 49 | 120 | 187 | 23 | 121 | 150 | 28 |
|  |  | 200 | 262 | 17 | 155 | - | - |
| ZAC | 24 | 120 | 30 | 55 | 72 | 224 | 10 |
|  |  | 200 | 42 | 43 | 92 | - | - |

**Table 2:** Elastic and inelastic mean free path values $\lambda_{elastic}$ and $\lambda_{inelastic}$ in aC and the ZAC-alloy for 120 and 200 keV electron energy. The scattering probabilities $P_{elastic}$ and $P_{inelastic}$ are calculated on the basis of Eq. (2). Measured values for $\lambda_{inelastic}$ are derived from the low-loss EELS spectra in Fig. 6.

the probability for elastic scattering, given in percent, can be calculated by

$$P_{elastic} = 100 \cdot (1 - \exp[-t/\lambda_{elastic}]) \quad (2)$$

Values of $P_{elastic}$ are listed in Table 2. The short elastic mean free path in the ZAC-alloy is not fully compensated by the reduced film thickness because $P_{elastic}$ in the ZAC-alloy is increased by a factor of ~ 2.5 compared to the aC-film. At 200 keV, 43 % of the electrons are elastically scattered in the ZAC-film as compared to 17 % in the aC-film. Elastic scattering in the PP-material causes a displacement of object information in the image plane. Elastically scattered electrons change their direction of propagation, which gives rise to a shadow image in some distance from the expected object position. However, the formation of shadow images can be avoided if the illumination is confined to the area of interest [18]. Plane-wave illumination conditions with a well-defined beam diameter can be obtained in transmission electron microscopes with a Koehler illumination system. Alternatively, plane-wave micro- or nano-beam illumination conditions can be set up in transmission electron microscopes with a three-condenser lens system.

Inelastic scattering in the PP-material then remains the only limiting factor, which leads to a loss of coherence. The resulting reduction of resolution seriously affects the usability of thin-film based PPs in high-resolution phase-contrast TEM applications, such as single particle analysis. Inelastic scattering comprises phonon and plasmon scattering as well as inner-shell ionization. The probability for inner-shell ionization is small compared to phonon and plasmon scattering, which are the dominating processes with energy losses of up to 30 eV. Phonon scattering predominantly occurs in large scattering angles. Phonon-scattered electrons in the object can be excluded from image formation by an appropriately sized objective aperture [31]. The inelastic mean free path $\lambda_{inelastic}$ can be measured from low-loss EELS spectra if the film thickness t is known [32]

$$\lambda_{inelastic} = \frac{t}{\ln\left(\frac{I_{ZL} + I_{LL}}{I_{ZL}}\right)} \quad (3)$$

The zero-loss and low-loss intensity are denoted by $I_{ZL}$ and $I_{LL}$. The intensities are obtained by integration of the zero-loss peak and the low-loss region up to 60 eV energy loss. Table 2 contains measured values for $\lambda_{inelastic}$ in aC and the ZAC-alloy derived from the low-loss EELS spectra in Fig. 6. For comparison, $\lambda_{inelastic}$ can be calculated by a semi-empirical expression [33]. Table 2 contains calculated values for $\lambda_{inelastic}$, which is considerably shorter in the ZAC-alloy than in aC. Experimental and calculated inelastic mean free path values in aC are in good agreement. However, a considerable deviation is found for the ZAC-alloy, where the measured inelastic mean free path of 224 nm is three times larger than the calculated value of 72 nm. Similar discrepancies were found for other metallic glasses like $Zr_{50.0}Al_{5.0}Cu_{45.0}$ with similar properties as the ZAC-alloy [34]. The discrepancy between calculated and measured $\lambda_{inelastic}$ values can be explained because the Malis formula is only valid for small collection angles [35] and electron energies below 100 keV [36]. The probability for inelastic scattering in the PP-material can be calculated according to Eq. (2) by replacing $\lambda_{elastic}$ with $\lambda_{inelastic}$.

Due to the extraordinary large inelastic mean free path in the ZAC-alloy, the inelastic scattering probability in the ZAC-film is only 10 % compared to 28 % in aC.

The small probability for plasmon scattering in the ZAC-alloy is also visualized in the power spectrum and the corresponding Thon-ring intensity profiles (Figs. 5b,c). Inelastic scattering causes a loss of coherence, which leads to an attenuation of the phase-contrast transfer function and a reduction of the Thon-ring intensity above the cut-on frequency. For the ZAC-alloy, the Thon-rings in regions below and above the cut-on frequency are of comparable intensity (Fig. 5b), which is a strong indication of the weak inelastic scattering probability. Moreover, the information limit is not seriously affected as shown in Fig. 5c. Both azimuthally averaged intensity profiles reveal distinct oscillations up to the same spatial frequencies. The effect of low-probability plasmon scattering is even more pronounced for Zernike PPs, where only 50 % of the HPP film thickness is required. In this case, an overall plasmon scattering probability of only 5 % is expected in the ZAC-film compared to 15 % in aC.

## 5. Summary

This work considers for the first time thin-film based PPs fabricated from metallic glass alloys. PPs of the zirconium-based ZAC-alloy were prepared and characterized by different TEM techniques. The experimental results show that metallic glasses are promising to improve the performance of thin-film based PPs in phase-contrast TEM. Besides its high electrical conductivity, the ZAC-alloy provides several beneficial properties:

- The ZAC-alloy forms an amorphous structure, which is well preserved even after several hours of intense electron-beam illumination.

- The measured MIP of the ZAC-alloy (21.6 V) is more than twice as high as the MIP of aC (8.9 V), which allows a substantial reduction of the film thickness.

- The probability for inelastic plasmon scattering in the ZAC-alloy is low due to the extraordinary large inelastic mean free path. Inelastic mean free path values were measured at 120 kV by evaluating the plasmon intensity in low-loss EELS spectra. For a 24 nm ZAC-film, the probability for inelastic scattering is only 10 % compared to 28 % in aC. This effect significantly improves contrast transfer at high spatial frequencies, which is required for high-resolution applications of PP-TEM.

The only unfavorable effect is oxide formation at the surface of the ZAC-film, which leads to electrostatic charging upon electron-beam illumination. The effect can be minimized by means of a thin aC-coating. Hence, at least for now, aC cannot be completely excluded from PP-fabrication. However, in contrast to pure aC-film based PPs, the charging of ZAC-film based PPs is reversible after a few hours as indicated by the disappearance of distortions in the Thon-ring system. Finally, it is mentioned that the group of metallic glasses includes a wide variety of different alloys, whose properties might be even better suited for PP-application (e.g. palladium-based systems).

**Acknowledgement**

The authors acknowledge funding of this project by the German Research Foundation (Deutsche Forschungsgemeinschaft) under contract Ge 841/26.